\documentclass{emulateapj}

\newcommand {\be} {\begin{equation}}
\newcommand {\ee} {\end{equation}}

\def\gtorder{\mathrel{\raise.3ex\hbox{$>$}\mkern-14mu
    \lower0.6ex\hbox{$\sim$}}}
\def\ltorder{\mathrel{\raise.3ex\hbox{$<$}\mkern-14mu
    \lower0.6ex\hbox{$\sim$}}}

\shorttitle{Magnetic flux paradigm for radio-loudness}

\shortauthors{Sikora \& Begelman}

\begin{document}

\title{Magnetic flux paradigm for radio-loudness of AGN}
\author{Marek Sikora$^{1}$ and Mitchell C. Begelman$^{2,3}$} 
\affil{$^{1}$
Copernicus Astronomical Center, Polish Academy of Sciences, ul. Bartycka 18, 00-716 Warsaw, Poland}
\affil{$^{2}$
JILA, University of Colorado and National Institute of Standards and Technology,
440 UCB, Boulder, CO 80309, USA}
\affil{$^3$
Department of Astrophysical and Planetary Sciences, University of Colorado, 
391 UCB, Boulder, CO 80309, USA}
\email{sikora@camk.edu.pl; mitch@jila.colorado.edu}

\begin{abstract}

We argue that the magnetic flux threading the black hole, rather than black hole spin or Eddington ratio, is the dominant factor in launching powerful jets and thus determining the radio loudness of active galactic nuclei (AGN).  Most AGN are radio quiet because the thin accretion disks that feed them are inefficient in depositing magnetic flux close to the black hole.  Flux accumulation is more likely to occur during a hot accretion (or thick disk) phase, and we argue that radio-loud quasars and strong emission-line radio galaxies occur only when a massive, cold accretion event follows an episode of hot accretion.  Such an event might be triggered by the merger of a giant elliptical galaxy with a disk galaxy.  

This picture supports the idea that flux accumulation can lead to the formation of a so-called magnetically choked accretion flow (MCAF).  The large observed range in radio loudness reflects not only the magnitude of the flux pressed against the black hole, but also the decrease in UV flux from the disk, due to its disruption by the ``magnetosphere" associated with the accumulated flux.  While the strongest jets result from the secular accumulation of flux, moderate jet activity can also be triggered by fluctuations in the magnetic flux deposited by turbulent, hot inner regions of otherwise thin accretion disks, or by the dissipation of turbulent fields in accretion disk coronae.  These processes could be responsible for jet production in Seyferts and low-luminosity AGN, as well as jets associated with X-ray binaries. 

\end{abstract}


\section{Introduction}

Half a century after the discovery of quasars, we are still struggling to understand the enormous diversity of their jet activities.  In most quasars the jet power, $P_j$, is a factor $\sim 10^4$ times smaller than the accretion power, $P_{\rm acc}$, while others have jet powers covering the range extending up to $P_j \sim P_{\rm acc}= \epsilon \dot M c^2$ (Rawlings \& Saunders 1991; Ghisellini et al.~2010; Fernandes et al.~2011; Punsly 2011), where $\dot M$ is the accretion rate and $\epsilon$ the accretion efficiency.  Such a broad range of jet production efficiencies, $P_j/P_{\rm acc}$, can be explained via the Blandford-Znajek mechanism for powering jets (Blandford \& Znajek 1977) if there is a sufficiently large spread of black hole spins and/or magnetic fluxes that thread the holes.

The broad range of jet production efficiencies is seen not only in quasars, but also in AGN with luminosities much smaller than the Eddington limit (Sikora et al.~2007, and references therein). Radio-selected AGN have been found to produce jets with efficiency typically 3 orders of magnitude larger than radio-detected, optically-selected AGN.  To explain such a difference solely in terms of black hole spin (i.e., according to the ``spin paradigm") requires dimensionless spin parameters spread over the range $a \sim 0.03 - 1$.  But according to numerical simulations of the cosmological evolution of black hole spins, such a range is rather difficult to reproduce (Volonteri et al.~2007, 2012).  A phenomenological ``spin-accretion" paradigm, adopting the Eddington ratio as a second parameter, also fails to explain the observed range of radio loudness (Sikora et al.~2007).  This suggests that the main parameter driving the diversity of jet production efficiencies is the magnetic flux (Sikora et al.~2013).

In the case of the most powerful jets, with $P_j \sim P_{\rm acc}$, the required magnetic flux is too large to be confined by the ``static" pressure of the accretion disk, even for maximal black hole spins (Moderski \& Sikora 1996; Ghosh \& Abramowicz 1997). However, this limitation can be overcome if very large magnetic fluxes can be forced onto the black hole by the ram pressure of the magnetically affected accretion flow.  Such a scenario is predicted by models that involve the central accumulation of large magnetic fluxes and the formation of magnetically arrested/choked accretion flows (MCAFs: Narayan et al.~2003; Reynolds et al.~2006; McKinney et al.~2012).

The idea that a large net magnetic flux can be accumulated around a black hole was already in place in the 1970s (Bisnovatyi-Kogan \& Ruzmaikin 1976).  However as Lubow et al.~(1994) pointed out, large-scale poloidal magnetic fields cannot be dragged to the center by standard, geometrically thin accretion disks, because of the outward diffusion of magnetic field in the turbulence triggered by the magneto-rotational instability (MRI).  Noting that no such limitation applies in the case of geometrically thick accretion flows (Lubow et al.~1994; Livio et al.~1999; Cao 2011; Guilet \& Ogilvie 2012), Sikora et al.~(2013) suggested that in radio-loud AGN the large magnetic fluxes were accumulated during a hot, low-accretion-rate phase prior to the current, cold accretion event.  We explore this idea below and try to verify whether magnetic flux can be the main parameter responsible for the observed spread in radio loudness.

Our paper is organized as follows. In \S2 we derive the conditions that must be satisfied by the hot and cold accretion rates and by the initial inner radius of the cold accretion disk (within which the flux is trapped), in order to initiate a magnetically-choked accretion flow (MCAF).  We next (\S3) estimate the resulting efficiency of black-hole rotational energy extraction via the Blandford-Znajek mechanism, and investigate whether the Blandford-Znajek mechanism can explain the observed broad range and distribution of radio loudness of AGN (\S4).  In \S5 we speculate about why the flux accumulation in most AGN is inefficient, discuss possible cosmological scenarios for black hole evolution that may lead to the observed radio-demographics of AGN, and suggest additional observational consequences.

\section{Magnetic flux accumulation}

Suppose a ``hot" (geometrically thick) accretion flow, with mass flux $\dot M_h$, drags a net poloidal magnetic flux $\Phi_{\rm h, tot}$ into the central region of an AGN.  Provided that this flux is larger than the maximum that can be confined on the BH by the ram pressure of the accreting plasma,
\be \Phi_{\rm h,tot} > \Phi_{\rm BH,max}(\dot M_h) = \phi (\dot M_h c r_g^2)^{1/2} \, , \ee
such an accumulation will result in the formation of a ``magnetosphere"
extending out to a radius
\be r_m(\dot M_h) \simeq \left(\frac{\Phi_{\rm h, tot}}{\Phi_{\rm BH,max}(\dot M_h)}\right)^{4/3}
\, r_g \, , \ee
(Narayan et al.~2003), where $r_g = GM_{\rm BH}/c^2$ is the gravitational radius and $\phi$ is a dimensionless factor which, according to numerical simulations by McKinney et al.~(2012), is typically of order 50.  The amount of magnetic flux enclosed within a radius $r<r_m(\dot M_h)$ is
\be \Phi_h(r) \simeq  \Phi_{BH,max}(\dot M_h) \left(r \over r_g \right)^{3/4} \, .
\ee

If the hot accretion phase is followed by a cold accretion event initiated by the formation of a cold disk at $r_d < r_m(\dot M_h)$, that disk will trap the flux  
\be \Phi_d = \Phi_h(r=r_d) \ee
and squeeze it to a radius
\be r_m(\dot M_d) \simeq
\left(\frac{\Phi_d}{\Phi_{\rm BH,max}(\dot M_d)}\right)^{4/3} \, r_g
\simeq \left(\dot M_h \over \dot M_d \right)^{2/3} r_d \, , \ee
or entirely enclose it on the black hole depending on whether $\Phi_d$ is larger or smaller than $\Phi_{\rm BH,max}(\dot M_d) \sim  \phi (\dot M_d c r_g^2)^{1/2}$.  Thus, a magnetosphere with $r_m(\dot M_d) > r_g$ is formed and the MCAF scenario proceeds if cold accretion at a rate $\dot M_d$ was preceded by hot accretion (satisfying condition [1]) at a rate 
\be \dot M_h > (r_d/r_g)^{-3/2} \dot M_d \, . \label{mcaf} \ee

\section{Jet production efficiency} 

If the condition given by equation (\ref{mcaf}) is satisfied, then $\Phi_d >\Phi_{\rm BH,max}(\dot M_d)$ and the rate of energy extraction from the rotating black hole via the Blandford-Znajek mechanism is
\begin{eqnarray}
P_{BZ}^{(MCAF)} & \simeq & 
4 \times 10^{-3} \Phi_{\rm BH,max}^2(\dot M_d)\frac{\Omega_{\rm BH}^2}{c} f_a(\Omega_{BH})
\nonumber \\
& = & 10 (\phi/50) x_a^2 f_a(x_a) \dot M_d c^2 \, , \end{eqnarray} 
where $\Omega_{\rm BH}$ is the angular velocity of the black hole, 
\be x_a = r_g \Omega_{BH}/c = [2(1+\sqrt{1-a^2})]^{-1} \, a \, , \ee 
$f_a(x_a) \simeq 1+1.4x_a^2-9.2x_a^4$, and $a$ is the dimensionless angular momentum parameter (Tchekhovskoy et al. 2010).  Although energy is also extracted from the magnetospheric region outside the event horizon, the latter can dominate the jet power only for very slowly rotating holes ($a<0.1$: McKinney et al.~2012).  Assuming $a>0.1$, we find that the efficiency of jet production in the MCAF scenario is 
\be \eta_j \equiv P_j/\dot M_d c^2 \simeq 10 (\phi/50) x_a^2 f_a(x_a) \, . 
\label{etaj} \ee
This gives $\eta_{j} \simeq 1.9 (\phi/50)$ for $a=1$ and $\eta_{j} \simeq 0.0063$ for $a=0.1$, and therefore the jet efficiency varies by a factor $\simeq 300$ over the spin range $0.1 < a < 1.0$.  For the spin range $0.4 < a < 0.9$, however, where most black holes are expected to lie, the efficiency varies by less than a factor 10.
 
\section{Radio loudness}

According to the analysis by Willott et al.~(1999), the radio luminosity of a typical extended radio source is approximately proportional to the jet power.  Adopting the modifications and approximations discussed by Sikora et al.~(2013), we use $P_j  \sim 10^2 \nu_{1.4} L_{\nu_{1.4}} (f/3)^{3/2}$, where $\nu_{1.4} = 1.4$ GHz and $f$ is a factor that is predicted to be in the range $1 < f < 20$ (Willott et al.~1999).  Then, using the standard definition of the radio-loudness, ${\cal R} = L_{\nu_5}/L_{\nu_B}$, where $\nu_5 = 5$ GHz and 
$\nu_B = 6.8 \times 10^{14}$ Hz (Kellermann et al.~1989), we obtain that the radio loudness predicted by the MCAF model is 
\be 
{\cal R} \equiv \left(\frac {\nu_B}{\nu_5}\right) \,
\frac{\nu_5 L_{\nu_5}}{\nu_B L_{\nu_B}}   \simeq 
10^4 \frac{P_j}{(f/3)^{3/2}P_d} \simeq 
10^4 \frac{\eta_j}{\epsilon_d (f/3)^{3/2}} \ee
where $\epsilon_d \equiv P_d/\dot M_d c^2$ is the radiative efficiency of the accretion disk, $ P_d \sim 10 \nu_BL_{\nu_B}$, and $L_{\nu_5}$ is related  to $L_{\nu_{1.4}}$ assuming a radio spectral index $\alpha_r = 0.8$.

In the MCAF scenario the outer, radiative viscous disk is truncated at $r_m(\dot M_d)$.  At smaller radii, the angular momentum of the accreting material is transferred to the magnetic field via interchange instabilities, allowing accretion to occur without substantial heating and radiative losses. We therefore estimate the radiative efficiency to be 
\be
 \epsilon_d  \sim \frac{r_g}{r_m(\dot M_d)} \simeq 
\left(\frac{\Phi_{\rm BH,max}(\dot M_d)}{\Phi_d}\right)^{4/3}  \simeq 
\frac{r_g}{r_d} \, \left(\frac{\dot M_d}{\dot M_h}\right)^{2/3} \, ,
\label{epsilon1} \ee
and then
\be {\cal R} \simeq 10^4 \, \frac{\eta_j}{(f/10)^{3/2}}\, \frac{r_d}{r_g} \, 
\left(\frac{\dot M_h}{\dot M_d}\right)^{2/3} \,  . \label{loudness1}\ee
Replacing $\dot M_d$ by the Eddington-ratio parameter, $\lambda \equiv P_d/L_{\rm Edd} = \epsilon_d \dot M_d c^2 / L_{\rm Edd}$, in equations (\ref{epsilon1}) and (\ref{loudness1}) gives
\be \epsilon_d = \left(\frac{r_g}{r_d}\right)^{3/5}
\, \left(\frac{\lambda}{\dot m_h}\right)^{2/5} \sim 
\left(\frac{\lambda}{\lambda_{\rm max}^{MCAF}}\right)^{2/5}   \, \ee
and 
\be {\cal R} \simeq 10^4 \, \frac{\eta_j}{(f/3)^{3/2}} \,
\left(\frac{\lambda}{\lambda_{\rm max}^{MCAF}}\right)^{-2/5} \, , 
\label{lambda} \ee
where $\dot m_h \equiv \dot M_h c^2/L_{\rm Edd}$, and 
\be
\lambda_{\rm max}^{MCAF} = \left(\frac{r_d}{r_g}\right)^{3/2} \, \dot m_h \,  \label{lambdamax}
 \ee
is the maximal Eddington-ratio achievable in the MCAF for given values of $\dot m_h$ and $r_d$ (see equation \ref{mcaf}).

\subsection{Quasars}

Studies of nearby quasars indicate that their radio-loudness distribution is bimodal, with the majority of quasars narrowly clustered around ${\cal R} \sim 1$ and others forming a tail extending up to ${\cal R} \sim 10^4$ (Kimball et al.~2011; Balokovi\'c et al.~2012; Singal et al. 2012).  While radio emission of quasars with ${\cal R} < 10$ is likely to be associated with stellar processes in star formation regions, radio emission of quasars with ${\cal R} > 10$  must involve dissipation of jet energy.  Values of ${\cal R} > 100$ are achievable for jets produced via the Blandford-Znajek mechanism in the MCAF scenario.  The radio-loudness of such ``MCAF"--quasars is expected to cover a range from ${\cal R} \sim$ tens up to ${\cal R} \sim 10^4$, with a spread determined by the range of black hole spins and of the Eddington-ratio parameter $\lambda$ (see equations [\ref{etaj}] and [\ref{lambda}]). 

Observations show a bottom-heavy distribution of quasar radio-loudness, with only a few percent of quasars having ${\cal R}>100$ (de Vries et al.~2006; Lu et al.~2007; Balokovi\'c et al.~2012).  According to the magnetic flux paradigm, this may indicate a bottom-heavy distribution of the hot accretion rate $\dot M_h$ (or the absence of a hot accretion phase altogether), leading to a small fraction of quasars that satisfy the MCAF condition (\ref{mcaf}).  Alternatively, some systems could fail to meet the flux threshold condition (1), perhaps because the hot accretion phase is too brief or the external field too disordered, implying that the accumulated magnetic flux is too weak to support the magnetosphere and is entirely crushed into the hole by the disk flow.  These ``MCAF--failed" quasars produce less powerful jets.  Such jets are likely to be efficiently decelerated within the galactic optical cores and most of them become subsonic with Fanaroff-Riley type I or other centrally dominated morphologies (Komissarov 1994; Bicknell 1995; Heywood et al.~2007). This can explain why radio emission in quasars with ${\cal R} < 100$ is dominated by compact sources (Lu et al.~2007).  

Radio emission at a level corresponding to ${\cal R} \ltorder 100$ could also be contributed by jets associated with fluctuating magnetic fields. In the case of quasars, such fields would probably arise in a corona above the thin accretion disk, and the jet could be propelled thermally by efficient heating of the coronal plasma due to magnetic reconnection (Heinz \& Begelman 2000).
In sources where the inner region of the accretion flow becomes hot or geometrically thick for any reason, large magnetic field fluctuations could develop with coherence length $\sim r$ and strength of order the ram pressure.  These flows could deposit transient flux onto the black hole, leading to intermittent jet production.  We tentatively associate these stochastic jets with the radio activity detected in radio-intermediate quasars, low-luminosity AGN, and X-ray binaries. 
 
\subsection{Strong-line radio galaxies (SLRGs)}

Strong,  broad-line and narrow-line radio galaxies (BLRGs and NLRGs, respectively) form together with radio-selected quasars an Eddington-ratio sequence that extends down to $\lambda \sim 10^{-4}$ (Sikora et al.~2007, 2012). Sikora et al.~(2013) showed that their radio loudness anticorrelates with the Eddington ratio, in accordance with  the MCAF model prediction (see equation [\ref{lambda}]). 

\subsection{Seyferts}   

Seyferts dominate the population of radio-quiet AGN.  Though covering a similar Eddington-ratio range as SLRGs, they are 2--4 orders less radio-loud, with ${\cal R}\sim 1$ like radio-quiet quasars (Sikora et al.~2007).  According to the magnetic flux paradigm, this implies very inefficient flux accumulation prior to the start of the Seyfert activity phase.  This inefficiency can be attributed to the lack of a hot accretion pre-phase.  In the case of objects hosted by disk galaxies, this conjecture is supported by studies of the nuclei of disk galaxies at epochs of very low accretion rates.  Broad-band spectra of such low-luminosity AGN (LLAGN) indicate that at least the outer portions of their accretion disks are cold and geometrically thin (Nemmen et al.~2012; Yu et al.~2011). 
 
Nevertheless, LLAGNs are found to be moderately radio-loud (Ho 2002).  Their radio activity, like that in low/hard states of XRBs, is presumably associated with a presence of a hot, radiatively inefficient accretion zone extending out to some distance from the black hole.  As discussed in \S 4.1, geometrically thick flows enable a significant poloidal magnetic field to impinge on the black hole and thus generate jets or winds (Livio et al.~1999).  
 
\subsection{X-ray binaries (XRBs)}

The MCAF scenario was suggested by Igumenshchev (2009) to explain jet activity and state transitions in XRBs.  However, noting that the outer portions of the accretion disks in such objects are geometrically thin, it is not clear how magnetic flux can be advected efficiently to the vicinity of the black hole.  An alternative scenario may involve fluctuating magnetic fields, as mentioned earlier (\S 4.1). In low/hard states this could involve a central hot, geometrically thick accreting region with large--scale fluctuating poloidal magnetic fields, while in high/hard states the dominant mechanism could be thermal propulsion following dissipation of tangled magnetic fields in the disk corona.

\section{Evolutionary and observational considerations}
 
The proposed flux paradigm explains the relative rarity of radio-loud AGN because it requires a special sequence of events to occur in order to trigger powerful jet activity.  This contrasts with the spin paradigm, which appeals to a bottom-heavy distribution of black-hole spins that seems implausible in the light of recent evolutionary modeling. 

The flux paradigm fits well with the observationally supported idea that the most luminous AGN  are triggered by mergers (Hirschmann et al.~2012; Treister et al.~2012; Ramos Almeida et al.~2012).  To produce a radio-loud quasar or SLRG, a disk galaxy would presumably merge with a giant elliptical where hot accretion from the interstellar medium has been going on for some time.  The precondition of hot accretion is actually a dual requirement, because the hot accreting gas must also be carrying sufficient magnetic flux.  Thus we conjecture that the hot interstellar medium in at least some ellipticals contains a relatively coherent magnetic field that can be dragged inward.  While the origin of large-scale magnetic fields in galaxies is still under debate, it is plausible that these fields are seeded by stellar processes, through winds and supernova explosions, which suggests that the occurrence of powerful jets in AGN should be correlated with the cosmic history of star formation, separately from any generic correlation that links the growth of supermassive black holes to star formation.  In the central galaxies of cool-core clusters and groups, jet production is ubiquitous, suggesting that the gas in these relatively dense hot environments generally carries net magnetic flux (Burns 1990; Hardcastle et al.~2007; Tasse et al.~2008; Dunn et al.~2010). 
   
The same cold accretion flow responsible for amplifying radio-loudness could also serve as the circumbinary disk needed to drive the black holes in merging galaxies through the ``final parsec" toward merger (Begelman et al.~1980; Cuadra et al.~2009; Dotti et al.~2012, and references therein).  In this case, we could relate the radius within which the cold disk gathers up the trapped flux, $r_d$, to the circumbinary radius, which could be $\sim 10^3-10^4 r_g$, or larger.

To explain powerful jets, the MCAF scenario requires not only a large magnetic flux but also a substantial black hole spin.  Because accreting matter gives up its angular momentum to the dynamically dominant magnetic fields as it falls through the magnetosphere, it reaches the black hole with almost no angular momentum.  As a result, angular momentum extracted from the black hole during the MCAF phase is not replenished by the accreted matter, and the black hole spin decreases.  Thus, an additional requirement for a high level of radio loudness is that the black hole should have a large spin prior to the start of the MCAF phase.  

Considering the above requirements for producing radio-loud quasars, one might envisage two evolutionary tracks leading to radio-quiet quasars.  One track would involve the merger of two disk galaxies, where magnetic flux has never been collected through an extended period of hot accretion.  Another track would represent the last phases of MCAF activity where the black hole has been spun-down to extremely small values of $a$.  Noting, however, that significant reduction of a black hole's spin by zero-angular-momentum inflow requires its mass to be approximately doubled, and that such a doubling is not achievable during the typical lifetime of a classical double radio-source, $t_{FRII} \sim 3 \times 10^7$years (O'Dea et al.~2009), we regard the latter option as rather unrealistic unless $\dot M_d > \dot M_{Edd}$. Furthermore, an evolutionary connection between radio-loud and radio-quiet AGN seems to be excluded by observations showing that the hosts of radio-loud AGN are on average located in denser environments than those of radio-quiet AGN (Shen et al.~2009; Donoso et al.~2010; Sabater et al.~2012). 

We have focused mainly on the possible role of MCAFs in explaining the most luminous, radio-loud AGN, but emphasize that other processes may trigger jets as well, albeit giving a lower level of radio-loudness.  We suggest that any hot or geometrically thick inner region of an accretion flow can develop magnetic field fluctuations of large enough spatial coherence and strength  to produce an intermittent jet.  The magnetic fields in these jets would undergo frequent reversals of polarity, possibly leading to enhanced dissipation through reconnection across the current sheets (Sikora et al.~2003) and resulting radiative signatures that could be used to distinguish them observationally from the jets produced by coherent magnetic flux in MCAFs.  We tentatively associate these stochastic jets with the radio activity detected in radio-intermediate quasars, low-luminosity AGN, and X-ray binaries in low/hard states.  In quasars, Seyferts and XRBs in high/hard states, where thin disk flows probably extend all the way to the black hole, radio-emitting jets or winds could be produced through the dissipation of fluctuating coronal magnetic fields.    

As for testing the MCAF picture directly, we note that a cold accretion flow, disrupted by magnetic stresses inside the magnetospheric radius $r_m$, may have a unique structure that would allow it to be distinguished from a hot inner accretion flow or a thin disk enveloped by a hot corona.  We expect the MCAF to consist of cold blobs or filaments dropping through the magnetosphere via interchange instabilities.  These blobs might reprocess the surrounding disk emission more effectively than a thick disk or corona, and also provide more optical depth across the inner region surrounding the black hole. Furthermore,the power-density spectrum of X-ray variations should have a high-frequency break at lower frequency in MCAFs than in viscous accretion flows.  The best objects to look for signatures of MCAFs would be those with the most extended magnetospheres, primarily radio galaxies, both strong-line and weak-line.  

\section*{Acknowledgments}

We acknowledge support from the Polish Ministry of Science and Higher Education through the grant NCN DEC-2011/01/B/ST9/04845, from the National Science Foundation through grant AST-0907872, and from NASA's Astrophysics Theory Program through grant NNX09AG02G.  MCB thanks the Copernicus Astronomical Center for its hospitality during the early stages of this project.

\end{document}